\documentclass[iop]{emulateapj}

\usepackage{verbatim}

\newcommand{\zwindow}{$0.25<z<3$}
\newcommand{\zcosmos}{$0.25<z<1$}
\newcommand{\zuds}{$1<z<3$}
\newcommand{\mcut}{10.5}
\newcommand{\msun}{$M_{\odot}$}
\newcommand{\masslimit}{$M>10^{\mcut}~M_{\odot}$}
\newcommand{\ncum}{$n_{\rm c}=1.4\times 10^{-4}$~Mpc$^{-3}$}

\begin{document}

\accepted{for publication in ApJ}

\title{{\em HST}/WFC3 Confirmation of the Inside-Out Growth of Massive Galaxies at $0<\lowercase{z}<2$ and Identification of their Star Forming Progenitors at $\lowercase{z} \sim 3$\altaffilmark{*}}

\author{Shannon G. Patel$^1$, Pieter G. van Dokkum$^2$, Marijn Franx$^1$, Ryan F. Quadri$^3$, Adam Muzzin$^1$, Danilo Marchesini$^4$, Rik J. Williams$^3$, Bradford P. Holden$^5$, and Mauro Stefanon$^{6,7}$}

\affiliation{$^1$Leiden Observatory, Leiden University, P.O. Box 9513, NL-2300 AA Leiden, Netherlands; patel@strw.leidenuniv.nl\\
$^2$Department of Astronomy, Yale University, New Haven, CT 06520-8101, USA\\
$^3$Observatories of the Carnegie Institution of Washington, Pasadena, CA 91101, USA\\
$^4$Department of Physics and Astronomy, Tufts University, Medford, MA 02155, USA\\
$^5$UCO/Lick Observatory, University of California, Santa Cruz, CA 95064, USA\\
$^6$Observatori Astron\`{o}mic de la Universitat de Val\`{e}ncia, 46980 Paterna, Val\`{e}ncia, Spain\\
$^7$Physics and Astronomy Department, University of Kansas, Lawrence, KS 66045-7582, USA}

\altaffiltext{*}{Based on observations made with the NASA/ESA {\em Hubble Space Telescope}, obtained at the Space Telescope Science Institute. STScI is operated by the Association of Universities for Research in Astronomy, Inc. under NASA contract NAS 5-26555.}

\begin{abstract}
We study the structural evolution of massive galaxies by linking progenitors and descendants at a constant cumulative number density of \ncum\ to $z \sim 3$.  Structural parameters were measured by fitting S\'{e}rsic profiles to high-resolution CANDELS {\em HST} WFC3 $J_{125}$ and $H_{160}$ imaging in the UKIDSS-UDS at \zuds\ and ACS $I_{814}$ imaging in COSMOS at \zcosmos.  At a given redshift, we selected the {\em HST} band that most closely samples a common rest-frame wavelength so as to minimize systematics from color gradients in galaxies.  At fixed $n_{\rm c}$, galaxies grow in stellar mass by a factor of $\sim 3$ from $z \sim 3$ to $z \sim 0$.  The size evolution is complex: galaxies appear roughly constant in size from $z \sim 3$ to $z \sim 2$ and then grow rapidly to lower redshifts.  The evolution in the surface mass density profiles indicates that most of the mass at $r<2$~kpc was in place by $z \sim 2$, and that most of the new mass growth occurred at larger radii.  This inside-out mass growth is therefore responsible for the larger sizes and higher S\'{e}rsic indices of the descendants toward low redshift.  At $z<2$, the effective radius evolves with the stellar mass as $r_e \propto M^{2.0}$, consistent with scenarios that find dissipationless minor mergers to be a key driver of size evolution.  The progenitors at $z \sim 3$ were likely star forming disks with $r_e \sim 2$~kpc, based on their low S\'{e}rsic index of $n \sim 1$, low median axis ratio of $b/a \sim 0.52$, and typical location in the star-forming region of the $U-V$ versus $V-J$ diagram.  By $z \sim 1.5$, many of these star-forming disks disappeared, giving rise to compact quiescent galaxies.  Toward lower redshifts, these galaxies continued to assemble mass at larger radii and became the local ellipticals that dominate the high-mass end of the mass function at the present epoch.
\end{abstract}


\section{Introduction}

Massive galaxies in the nearby universe are generally comprised of quiescent elliptical and S0 galaxies.  The formation of such massive galaxies has been an active area of study.  Recent observations suggest that the properties of these quiescent galaxies (QGs) were much different at earlier times.  For example, the sizes of QGs have been found to be much smaller, at fixed stellar mass, at high redshift \citep[e.g.,][]{daddi2005,zirm2007,toft2007,vandokkum2008b,cimatti2008b,vanderwel2008c,franx2008,williams2010,newman2012}.  This implies much higher mass densities within the effective radius for QGs at high redshift.  The size measurements are robust \citep{szomoru2010,szomoru2012} and the stellar mass measurements are in good agreement with dynamical mass estimates \citep{cenarro2009,cappellari2009b,vandokkum2009b,vandesande2011}, confirming the dense nature of QGs at high redshift.  Several mechanisms have been proposed to explain the growth in sizes among QGs \citep[see, e.g.,][]{hopkins2010b}.  Recent discussions have centered on the relative importance of major and minor dissipationless mergers with the latter favored to be the primary channel for size growth \citep[e.g.,][]{bournaud2007c,naab2009b,bezanson2009,hilz2013}.  While evidence is emerging that these compact QGs represent the cores of local ellipticals \citep{bezanson2009,hopkins2009j,vandokkum2010}, the progenitors of these compact QGs at even higher redshifts remain a mystery.

Selecting galaxy samples at or above a fixed stellar mass limit has provided important insight into the evolution in properties for such populations.  However, the connection to the evolution of a typical galaxy over cosmic time is not straightforward given that galaxies grow in stellar mass due to in situ star formation and merging: the progenitors of galaxies that lie just above a stellar mass limit at low redshift would not be counted in a census of high redshift galaxies above the same mass limit since they were likely to be less massive.  Measuring the structural evolution of a galaxy as it grows in mass therefore requires a method for linking its progenitors and descendants over cosmic time.  One such method involves selecting galaxies at a constant cumulative number density.  The basic principle behind this method is that the rank ordering of galaxy masses does not change drastically over time.  Therefore, if one selects the 10th most massive galaxy in a comoving volume at $z \sim 3$, it is still likely to be approximately the 10th most massive galaxy in that comoving volume at $z \sim 0$, but with a higher overall mass due to star formation and merging.  This technique presents a complementary approach to mass-selected studies.  A number density selection has been used in other recent works to study the structural properties \citep{vandokkum2010}, star formation histories \citep[SFHs;][]{papovich2011}, and mass growth \citep{brammer2011} of galaxies over cosmic time \citep[see also,][]{loeb2003b}.  The study by \citet{vandokkum2010} was carried out with ground-based imaging.  In this work, we build on the results in \citet{vandokkum2010} and utilize high-resolution {\em Hubble Space Telescope} ({\em HST}) imaging, allowing us to measure structural properties more accurately and to push to $z \sim 3$.  At these high redshifts, we also identify the progenitors of $\sim 2M^{\star}$ galaxies, which in the local universe are bulge-dominated, quiescent systems with large effective radii.

The layout of this paper is as follows.  In Section~\ref{sec_data}, we discuss the data used for the study.  In Section~\ref{sec_analysis}, we discuss the relevant derived quantities.  In Section~\ref{sec_cumnum}, we examine the structural assembly of a galaxy as it grows in time to become a $\sim 2 M^{\star}$ galaxy by $z \sim 0$.  Our results are further discussed in Section~\ref{sec_discussion} and we summarize our findings in Section~\ref{sec_summary}.  For completeness, we briefly overview the structural properties of QGs and star-forming galaxies (SFGs) above a constant stellar mass limit in the Appendix, as this provides an alternative view to the number density selection in the main part of the paper.

We assume a cosmology with $H_0 = 70$~km~s$^{-1}$~Mpc$^{-1}$, $\Omega_M = 0.30$, and $\Omega_{\Lambda} = 0.70$.  Stellar masses are based on a Chabrier initial mass function \citep[IMF;][]{chabrier2003}.  Adopting IMFs of different forms below $1$~\msun\ \citep[e.g.,][]{vandokkum2010c} would lead, to first order, to an overall scaling of the stellar masses and otherwise identical results.  All magnitudes are given in the AB system.

\section{Data} \label{sec_data}

\subsection{UDS-CANDELS: $1<z<3$}

We study the structural properties of massive galaxies at high redshift (\zuds) using a combination of data from the UKIRT Infrared Deep Sky Survey \citep[UKIDSS;][]{lawrence2007} and the Cosmic Assembly Near-infrared Deep Extragalactic Legacy Survey \citep[CANDELS;][]{grogin2011}.  We employ a multi-wavelength data set ($u^{*}BVRi^{\prime}z^{\prime}JHK$ and {\em Spitzer} IRAC 3.6~\micron\ and 4.5~\micron) in the Ultra-Deep Survey (UDS) field of UKIDSS.  This is one of the deepest wide-field near-IR data sets available, making it ideal for constructing samples of galaxies to relatively low stellar masses at high redshift.  The observations and data reduction are described in detail in \citet{williams2009,williams2010} and in \citet{quadri2012} and briefly summarized here.  The UKIDSS UDS DR8 data include $JHK$, which have 5$\sigma$ limiting depths in $D=1\farcs8$ apertures of 24.9, 24.1, and 24.5~AB mag, respectively.  The $BVRi^{\prime}z^{\prime}$ imaging were obtained as part of the Subaru-$XMM$ Deep Survey \citep[SXDS,][]{sekiguchi2004}.  The $u^{*}$ data were obtained with MEGACAM on CFHT (PI: O. Almaini).  The {\em Spitzer} IRAC 3.6~\micron\ and 4.5~\micron\ data were obtained as part of the {\em Spitzer}-UDS Survey (SpUDS; PI: J. Dunlop).  An updated catalog will be presented in detail in Williams et al. (2013, in preparation).  Objects were detected in the $K$-band and photometry carried out in the other bands with matched apertures.  The IRAC photometry was measured using the point spread function (PSF) convolution procedure of \citet{labbe2006}.  The IRAC bands in the UDS are crucial for determining photometric redshifts, stellar masses, and rest-frame optical/near-IR colors of galaxies at high redshift.  An analysis of simulated number counts indicates that the catalog is $>90\%$ complete at $K_{\rm tot}=24.0$~AB mag.

To measure structural parameters, we utilize {\em HST} imaging in the UDS that was acquired as part of CANDELS\footnote{http://candels.ucolick.org/data\_access/Latest\_Release.html} \citep{grogin2011}.  The observations and data reduction are presented in \citet{koekemoer2011}.  We employ $v$1.0 of the publicly available WFC3 $J_{125}$ and $H_{160}$ mosaics.  The PSF FWHM in these two bands are $0\farcs12$ and $0\farcs18$, respectively.  The CANDELS {\em HST} WFC3 data cover roughly $\sim 0.06$~deg$^2$ of the $\sim 0.65$~deg$^2$ UDS field with multi-wavelength coverage.

\subsection{COSMOS-ACS: $0.25<z<1$}

In order to assemble a large sample of massive galaxies at low redshift (\zcosmos), we use data in the COSMOS field \citep{scoville2007,scoville2007b}.  This $\sim 2$~deg$^2$ field benefits from multi-wavelength imaging spanning the UV to IR, including wide-field {\em HST} ACS coverage.  \citet{ilbert2009} assembled a photometric catalog in COSMOS incorporating CFHT $u^{*}$ and $K$, Subaru $BVgriz$ as well as 12 intermediate optical bands (IA427, IA464, IA484, IA505, IA527, IA574, IA624, IA679, IA709, IA738, IA767, IA827), UKIRT $J$, all four IRAC channels, and {\em GALEX} NUV and FUV.  Some of these data products are further described in \citet{capak2007}.  \citet{ilbert2010} used the photometry in COSMOS to derive stellar mass estimates, which could in principle be used in our analysis.  However, for consistency with the data preparation and stellar masses computed in the UDS (see Section~\ref{sec_analysis}), we use a $K_s$-selected catalog, which incorporates most of the data in COSMOS described above but reconstructed in a manner consistent with what was done in the UDS \citep[e.g.,][]{williams2009,quadri2012}.  A detailed account of the observations, data reduction, and catalog construction will be described in a future paper (Muzzin~et al. 2013, in preparation).  We briefly remark on the most relevant points of the catalog here.  Objects were detected in the $K_s$-band, which reached a $5\sigma$ detection limit of 23.85~AB~mag for a $D=2\arcsec$ aperture.  An analysis of simulated number counts indicates that the 90\% completeness is $K_{s, {\rm tot}}=23.5$~AB mag.

The primary purpose of using data in the COSMOS field is the wide-field {\em HST} ACS $I_{814}$ imaging, which we use to measure structural parameters (see Section~\ref{sec_analysis}).  In this work, we utilize the $v2.0$ ACS $I_{814}$ imaging \citep{koekemoer2007,massey2010}.  The typical PSF FWHM is $0\farcs1$.  The ACS imaging employed here covers roughly $\sim 1.3$~deg$^2$ of the COSMOS field.

\section{Analysis} \label{sec_analysis}

\subsection{Photometric Redshifts, Stellar Masses, and Rest-frame Colors}

While the data sets used in this work are assembled from two different fields, the analysis carried out on the data is uniform.  Photometric redshifts in both COSMOS and UDS were measured with EAZY \citep{brammer2008}.  In the UDS, a comparison to spectroscopic redshift measurements suggests that the photometric redshift uncertainties are $\sigma_z/(1+z) \sim 0.022$ at $1<z<1.5$.  At higher redshifts, spectroscopic samples are limited therefore making comparisons to the photometric redshifts difficult.  As a result, losses and contamination from catastrophic outliers remain unquantified.  However, at $z>1.5$, where the majority of our sample in the UDS lies, \citet{quadri2012} find that the photometric redshift uncertainties are larger based on differences in the photometric redshifts of close pairs \citep{quadri2010}.  SFGs in particular have larger photometric redshift uncertainties.  In COSMOS, a comparison to spectroscopic redshifts at $0.25<z<1$ indicates that the photometric redshift uncertainties are $\sigma_z/(1+z) \sim 0.01$.  Rest-frame $U-V$ and $V-J$ colors were also computed with EAZY.  

Stellar masses were computed with the spectral energy distribution (SED) fitting code FAST \citep{kriek2009} using exponentially declining SFHs.  \citet[][hereafter BC03]{bc03} stellar population synthesis (SPS) models with a Chabrier IMF were used in the SED fitting.  In the UDS, \citet{quadri2012} parameterize the stellar mass limit as a function of redshift as $M_{\rm lim}=9.4+1.2\ln(z)$.  Using the same technique to compute the mass limit as \citet{quadri2012}, but focusing solely on galaxies near $z=3$, we find that a mass limit of $M \sim 10^{10.6}$~\msun\ encompasses $\sim 95\%$ of galaxies.  We adopt this value as our stellar mass limit in the UDS at $z=3$.  As will be made clear in Section~\ref{sec_cumnum}, galaxies in our COSMOS sample at \zcosmos\ are well above the limiting stellar mass at $z=1$ of $M \sim 10^{9.4}$~\msun.

\subsection{Structural Parameters}

Structural parameters were obtained with GALFIT \citep{peng2002}, which provided measurements of S\'{e}rsic indices ($n$), effective radii ($r_e$), and axis ratios ($b/a$).  The effective radii reported here are circularized, $r_e=\sqrt{ab}$.  In order to carry out these measurements at the same rest-frame wavelengths, we use the {\em HST} imaging that is closest in rest wavelength to $\lambda_0 = 5160$~\AA.  This choice of $\lambda_0$ represents a tradeoff between ($1$) the desire to push to high redshifts, and ($2$) the need to probe rest-frame light as redward as possible from the 4000~\AA\ break, beyond which the SEDs of galaxies are generally smooth and serve as a closer tracer of the stellar mass.  As a result, in COSMOS, we use ACS $I_{814}$ imaging at $0.25<z<1$, while in the UDS we use WFC3 $J_{125}$ imaging at $1<z<1.76$ and WFC3 $H_{160}$ imaging at $1.76<z<3$.  For a given object, the nearest star is chosen to serve as the PSF model when running GALFIT.  Nearby objects were masked.  No constraints were placed on the range of S\'{e}rsic indices.  The axis ratios were constrained to have $0.1<b/a<1$.  The semi-major axis, $a$, was constrained to be smaller than the box size.  

Our results and conclusions are dependent on the quality of the structural parameters measured from GALFIT.  We ran simulations to test the reliability of these measurements at the highest redshifts, where objects become faint in the WFC3 $H_{160}$ imaging.  We created $10,000$ mock galaxies from a range of S\'{e}rsic models, varying $r_e$, $n$, $b/a$, and the magnitude.   These models were added to different regions of the WFC3 $H_{160}$ imaging that contained blank sky.  We then processed the mock images in the same manner as used to derive our measurements above.  The sample in our highest redshift bin ($2.5<z<3$), defined in Section~\ref{sec_cumnum}, reaches magnitudes of $H_{160} \sim 23.6$~AB mag and $24.3$~AB mag at the 50th and 90th percentile, respectively.  Our simulations show that we recover $r_e$, $n$, and $b/a$ to precisions of $\sim 11\%$, $5\%$, and $2\%$ for $H_{160} \sim 23.6$~AB mag and $\sim 23\%$, $12\%$, and $5\%$ for $H_{160} \sim 24.3$~AB mag.  Systematic offsets are $<1\%$ for all three parameters.  Effective radii as small as $0\farcs06$ (i.e., less than the FWHM/2), or $r_e \sim 0.5$~kpc at $2<z<3$, are recovered to similar precisions as noted above, consistent with other works \citep[e.g.,][]{newman2012}.  Our results and conclusions based on structural parameters measured from GALFIT are therefore not strongly biased at high redshift in any way.  

As one might expect at higher redshifts, the higher resolution {\em HST} imaging reveals that a small portion of the objects detected in the ground-based $K$-band imaging are comprised of two or more objects blended together.  We remove galaxies in the UDS from our analysis ($\lesssim 10\%$) if nearby objects contribute more than $10\%$ of the total flux within the $D=1\farcs8$ color aperture, as the photometric redshifts, rest-frame colors, and stellar masses are not as reliable.  We note that this procedure may introduce a small bias against galaxies in close pairs.

\section{The Assembly of Massive Galaxies} \label{sec_cumnum}

\begin{figure*}
\epsscale{1.2}
\plotone{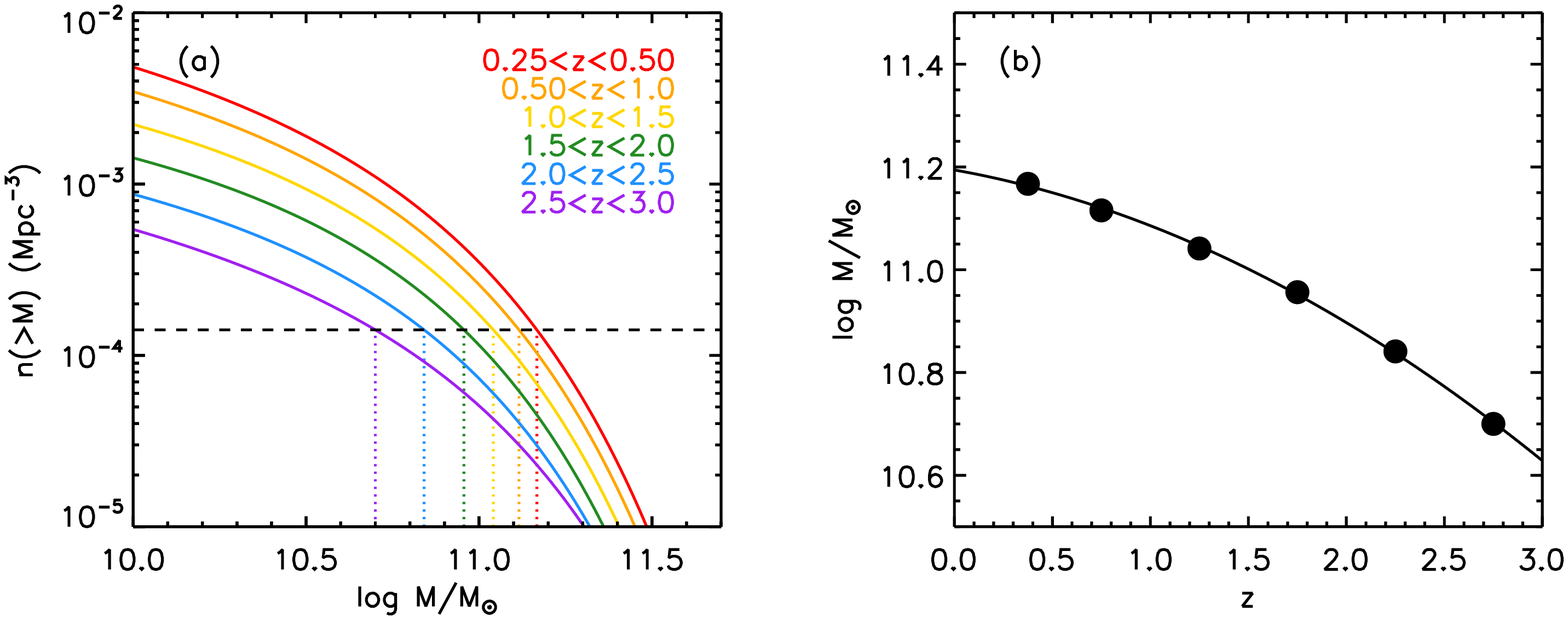}
\caption{(a) Cumulative number density of galaxies at a given stellar mass for different redshifts, derived from the mass functions of \citet{marchesini2009}.  At a fixed cumulative number density of \ncum\ (dashed black line) we determine the corresponding stellar mass for a given redshift bin (dotted vertical lines). (b)  Stellar mass vs. redshift for galaxies selected at \ncum.  The solid curve represents a second-order polynomial fit and is given by Equation~(\ref{eq_mass_redshift}).  A galaxy with a stellar mass of $M \approx 5\times10^{10}$~\msun\ at $z=2.75$ grows by a factor of $\sim 3$ in mass by $z=0.375$.  For a given redshift, we study the structural properties of galaxies at \ncum\ by selecting objects in a narrow mass bin around the predicted stellar mass from Equation~(\ref{eq_mass_redshift}).} \label{fig_cumnum_selection}
\end{figure*}

We define our sample in this section by selecting galaxies at a fixed cumulative number density, $n_{\rm c}$.  We then study the structural properties of galaxies in narrow mass bins at this value of $n_{\rm c}$ at different redshifts.  The analysis that follows is overall similar to that of \citet{vandokkum2010} but differs in a few aspects.  First, we study galaxies at a constant {\em cumulative} number density.  Second, our analysis extends to $z \sim 3$.  Finally, our work employs {\em HST} imaging as opposed to ground based, allowing us to more accurately measure structural properties, especially at high redshift.  The deep {\em HST} imaging also allows us to characterize the properties of individual galaxies as opposed to stacks, which were employed in \citet{vandokkum2010}.

\subsection{Selection at a Constant Cumulative Number Density}

\begin{figure*}
\epsscale{1.2}
\plotone{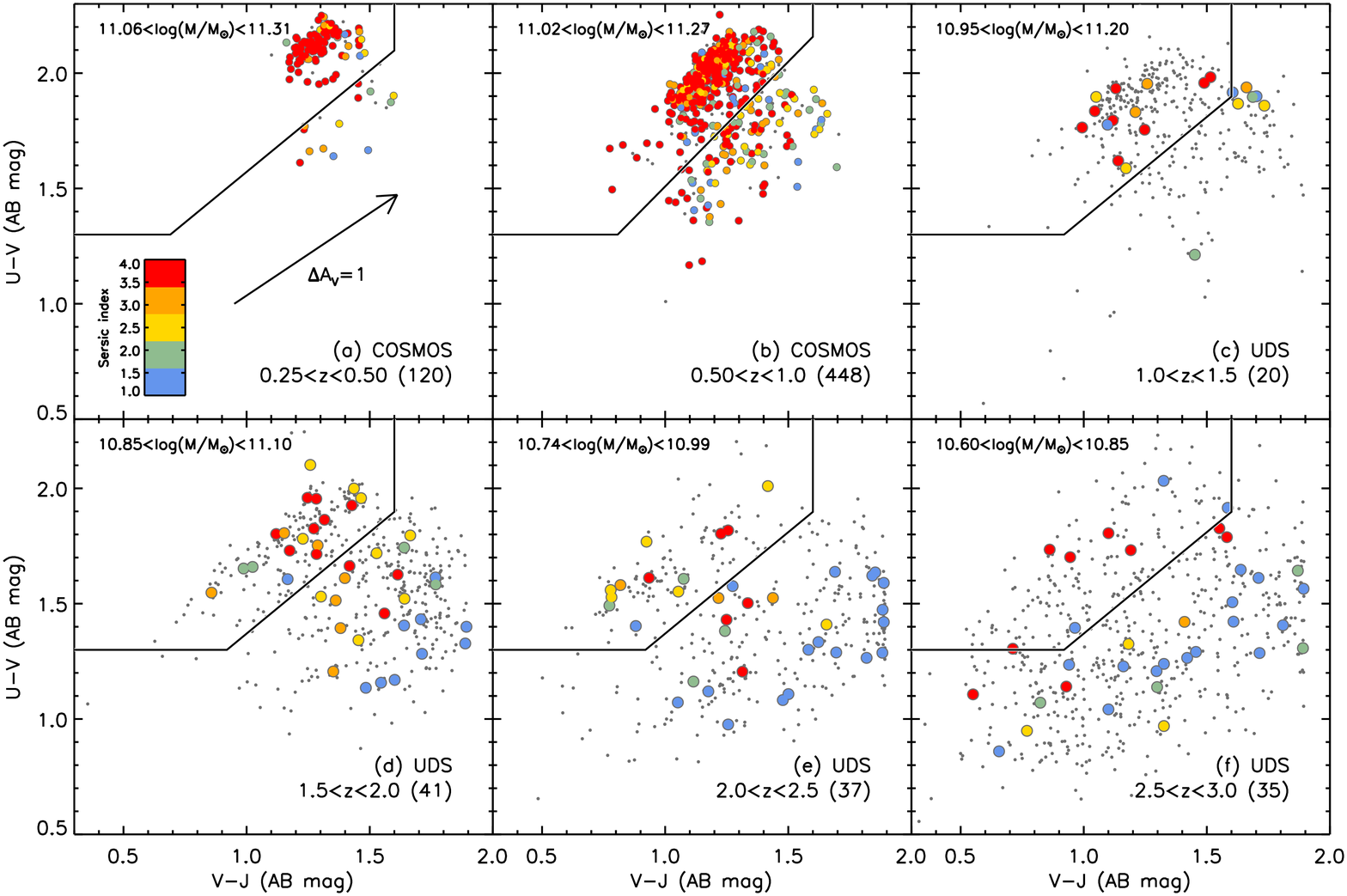}
\caption{Rest-frame $U-V$ vs. $V-J$ color for galaxies at a fixed cumulative number density, \ncum, in COSMOS (\zcosmos) and the UDS (\zuds).  Galaxies were selected in narrow mass bins at each redshift around the predicted stellar mass in Equation~(\ref{eq_mass_redshift}).  Note that the number density selection results in a selection of more massive galaxies toward low redshift (e.g., Figure~\ref{fig_cumnum_selection}).  The gray points indicate the parent sample of galaxies while objects color-coded by S\'{e}rsic index indicate the sample with {\em HST} imaging (note that the S\'{e}rsic indices were not bound to the range shown).  The sample size of galaxies with measured structural parameters is indicated in the bottom right of each panel.  The solid black line shows the division between QGs (top left) and SFGs (bottom right).  At $z>2$, the progenitors of $\sim 2M^{\star}$ galaxies were star-forming disks.  By $z \sim 1.5$, many of these compact star-forming disks have disappeared from the sample, while compact QGs have emerged.  At $z \sim 0.375$, the assembled $\sim 2M^{\star}$ galaxies resemble bulge-dominated, quiescent systems with large effective radii.} \label{fig_uvj_ncum}
\end{figure*}

Figure~\ref{fig_cumnum_selection}(a) shows the cumulative number density of galaxies at different redshifts.  These curves were derived using the mass functions of \citet{marchesini2009}, which were computed with data from several fields and as a result minimized uncertainties from cosmic variance.  In addition, \citet{marchesini2009} carefully accounted for completeness limits in stellar mass in deriving their mass functions.  We used best-fit Schechter parameters from Set~7 in that work, which used solar metallicity BC03 models with a Chabrier IMF and a \citet{calzetti2000} dust law to determine stellar masses, similar to the SFHs employed in this work.  We derived mass functions for our redshift bins by interpolating between the best-fit Schechter functions computed by \citet{marchesini2009}, which cover  $1.3<z<4$.  In carrying out the interpolation, we also included the \citet{cole2001} mass function at $z\sim 0.1$, as reported in \citet{marchesini2009} but scaled to a Chabrier IMF.  We then integrated these mass functions, $\Phi(M)$, to determine the cumulative number density of galaxies as a function of stellar mass at different redshifts:

\begin{eqnarray}
n(>M)=\int_{M}^{\infty} \Phi(M)dM
\end{eqnarray}

We chose a cumulative number density for our study of \ncum\ (dashed line) as this value represents the number density of galaxies with stellar masses slightly above the stellar mass limit at $z=3$ ($M \sim 10^{10.6}$~\msun).  We also note that the \citet{marchesini2009} mass functions are complete well below this value at these redshifts.  As seen in Figure~\ref{fig_cumnum_selection}(a), the selected value of $n_{\rm c}$ intersects the cumulative number density curves at lower redshifts at higher stellar masses, tracing out the mass growth at that particular number density.  The corresponding stellar masses at \ncum\ are shown as a function of redshift in Figure~\ref{fig_cumnum_selection}(b).  In order to quantify the redshift dependence we fit a second-order polynomial to these data points, resulting in the following relation between stellar mass and redshift:

\begin{eqnarray}
\log M_{n_{\rm c}}/M_{\odot}=11.19-0.068z-0.040z^2 \label{eq_mass_redshift}
\end{eqnarray}
The scatter about this relation is only $\sigma=0.0046$~dex, suggesting that this parameterization is adequate for the redshift range studied here.  Note that this scatter does not reflect the systematic uncertainty in measuring stellar masses of galaxies at high redshift, which can be substantial \citep[e.g.,][]{marchesini2009}.  Poisson uncertainties in the Schechter parameters computed by \citet{marchesini2009} propagate into an uncertainty in the derived stellar mass in Equation~(\ref{eq_mass_redshift}) at a given redshift of $\sim 0.10$~dex.  For the high-mass end at lower redshifts ($z \lesssim 1.5$), where cosmic variance is likely an important factor, an additional uncertainty of up to $\sim 0.10$~dex may be warranted.  For a given redshift, we study the properties of galaxies within a bin of size $\sim 0.3$~dex in stellar mass centered on the predicted mass from Equation~(\ref{eq_mass_redshift}).  The actual boundaries of the bin are adjusted such that the median mass is close to the value given by Equation~(\ref{eq_mass_redshift}).  Given the steepness of the mass function, in practice this results in selecting galaxies at $(\log M_{n_{\rm c}}/M_{\odot})_{-0.1}^{+0.15}$.  The bin size is broad enough to allow for robust measurements of median structural parameters.  Given the narrow redshift and mass bins employed, scattering of galaxies into and out of the sample due to photometric uncertainties is unavoidable and is a larger effect at higher redshifts (e.g., $z>2$) where Monte Carlo simulations of the photometry suggest uncertainties in the redshifts of $\sigma_z/(1+z) \sim 0.08$ and in the stellar masses of $\sim 0.17$~dex.  As a consequence, within a given redshift and mass bin at $z>2$, just under half of the original sample is recovered in our simulations while the remainder is made up of galaxies near the bin boundaries and therefore display similar properties to that of the original sample.  In order to avoid confusion, we emphasize that at a given redshift we are {\em not} selecting {\em all} galaxies with masses above the mass limit implied by the given value of $n_{\rm c}$, but instead, we are selecting galaxies in a narrow mass bin {\em at} the mass determined by $n_{\rm c}$.

Equation~(\ref{eq_mass_redshift}) indicates that galaxies at \ncum\ grow by a factor of $\sim 3$ from $z=2.75$ to $z \sim 0$, resulting in a galaxy at low redshift with a stellar mass of $M \sim 1.5 \times 10^{11}$~\msun\ (i.e., $\sim 2M^{\star}$).  From $z=2$ to $z=0.1$, Equation~(\ref{eq_mass_redshift}) predicts that the stellar mass grows by a factor of $\sim 2$, which is similar to what is found in \citet{vandokkum2010}.  We note that the stellar mass growth inferred from our purely observational motivated method (Equation~(\ref{eq_mass_redshift})) is less than what is predicted from abundance matching techniques \citep[e.g.,][]{conroy2009}, though the latter analysis is quite uncertain at $z>1$.  More recent efforts that combine dark matter merger trees with observational constraints indicate similar mass growth to what is found here at $z<3$ \citep{behroozi2012}.  At $z \sim 0$, the latter work indicates that a $\sim 2M^{\star}$ galaxy occupies a dark matter halo of mass $M \sim 4\times10^{13}$~\msun, which is typical of galaxy groups.

Both \citet{vandokkum2010} and \citet{papovich2011} show with simulations that in selecting galaxies at a fixed number density, the completeness fraction declines with cosmic time, meaning that some of the objects selected in a given number density bin at high redshift are no longer found in that bin at lower redshift.  Contaminants from other number density bins also enter the sample.  However, most of the contaminants scatter in from neighboring bins and likely display properties that are very similar to those of galaxies in the number density bin of interest.

Finally, we note that although we use mass functions from \citet{marchesini2009}, which are based on different data from what is employed here, we arrive at qualitative and quantitative conclusions that are quite similar to what was found in \citet{vandokkum2010}.  This suggests the systematic uncertainties as a result of this choice are minimal.

\subsection{Star Formation Properties} \label{sec_cumnum_uvj}

\begin{deluxetable*}{lcccccc}
\tablewidth{0pc}
\tablecolumns{7}
\tablecaption{Properties of Galaxies Selected at a Constant Cumulative Number Density of \ncum\  \label{table_ncum}}
\tablehead{
\colhead{Redshift} & \colhead{$N$\tablenotemark{a}} & \colhead{Mass\tablenotemark{b}} & \colhead{Quiescent\tablenotemark{c}} & \colhead{$r_e$\tablenotemark{d}} & \colhead{S\'{e}rsic\tablenotemark{e}} & \colhead{$b/a$\tablenotemark{f}} \\
\colhead{Range} &     \colhead{} &  \colhead{$\log M/M_{\odot}$} &      \colhead{Fraction} & \colhead{(kpc)} & \colhead{Index} & \colhead{}
}
\startdata
$0.25<z<0.5$ & 120  &  11.16  &  $0.89 \pm 0.03$  &  $6.1 \pm 0.3$  &  $4.6 \pm 0.2$  &  $0.74 \pm 0.01$ \\
$0.5<z<1$ & 448  &  11.12  &  $0.77 \pm 0.02$  &  $5.1 \pm 0.2$  &  $4.1 \pm 0.09$  &  $0.71 \pm 0.01$ \\
$1<z<1.5$ & 20  &  11.05  &  $0.65 \pm 0.1$  &  $3.2 \pm 0.5$  &  $3.2 \pm 0.5$  &  $0.64 \pm 0.05$ \\
$1.5<z<2$ & 41  &  10.95  &  $0.44 \pm 0.08$  &  $2.4 \pm 0.3$  &  $2.5 \pm 0.3$  &  $0.57 \pm 0.07$ \\
$2<z<2.5$ & 37  &  10.84  &  $0.32 \pm 0.08$  &  $2.3 \pm 0.3$  &  $1.8 \pm 0.4$  &  $0.68 \pm 0.06$ \\
$2.5<z<3$ & 35  &  10.70  &  $0.23 \pm 0.07$  &  $2.3 \pm 0.3$  &  $1.5 \pm 0.3$  &  $0.52 \pm 0.03$
\enddata
\tablenotetext{a}{Number of galaxies in the sample at the given redshift.}
\tablenotetext{b}{Stellar mass of galaxies at \ncum\ for a given redshift (see Equation~(\ref{eq_mass_redshift})).}
\tablenotetext{c}{Fraction of galaxies that are quiescent based on $UVJ$ selection.}
\tablenotetext{d}{Median effective radius for galaxies at $n_{\rm c}$.}
\tablenotetext{e}{Median S\'{e}rsic index for galaxies at $n_{\rm c}$.}
\tablenotetext{f}{Median axis ratio for galaxies at $n_{\rm c}$.}
\end{deluxetable*}

\begin{figure}
\epsscale{1.2}
\plotone{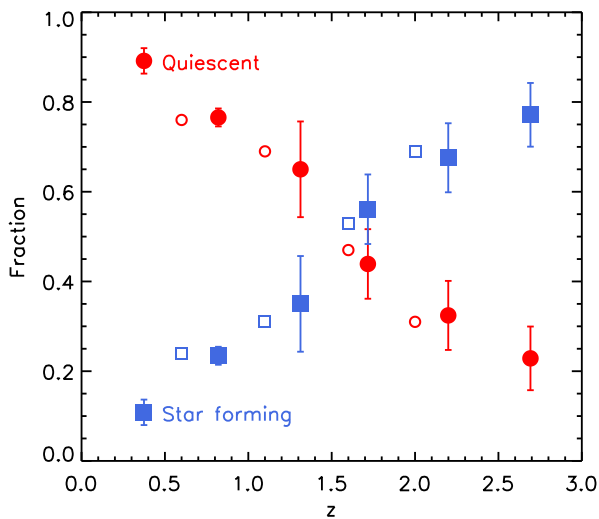}
\caption{Fraction of $UVJ$ classified quiescent galaxies (QGs, solid red circles) and star-forming galaxies (solid blue squares) vs. redshift for galaxies with measured structural parameters selected at a constant cumulative number density of \ncum.  The star-forming fraction is simply the complement of the quiescent fraction.  The $1\sigma$ error bars are computed assuming a binomial distribution.  The open symbols represent the values for the appropriate mass and redshift from \citet{brammer2011}.  The change in the proportion of QGs toward low redshift for galaxies at $n_{\rm c}$ is dramatic, increasing from $\sim 23\%$ at $z \sim 2.75$ to $\sim 89\%$ at $z \sim 0.375$.  At $z \sim 3$, most of the progenitors of massive galaxies were star forming.} \label{fig_cumnum_qg_frac}
\end{figure}

We first examine how the star formation properties of galaxies have evolved since $z \sim 3$.  As is well known, galaxies can be classified in two distinct categories: star forming and quiescent, at least out to $z \sim 3$ \citep{whitaker2011}.  Figure~\ref{fig_uvj_ncum} shows the rest-frame $U-V$ versus $V-J$ colors of galaxies in different redshift bins selected at \ncum.  This $UVJ$ diagram is commonly used to separate QGs from SFGs \citep[see, e.g.,][]{labbe2006,wuyts2007,williams2009,patel2011,patel2012}.  It is preferred over a color-magnitude or a color-mass selection because of its ability to separate red galaxies that are quiescent from reddened SFGs.  Shown in each redshift panel are galaxies within a narrow mass bin around the predicted stellar mass from Equation~(\ref{eq_mass_redshift}) for galaxies at $n_{\rm c}$.  The typical mass is therefore increasing toward low redshift.  The subset of galaxies with measured structural parameters from the {\em HST} imaging are color-coded according to their S\'{e}rsic index.  The \citet{williams2009} boundary distinguishing QGs (top left) from SFGs (bottom right) is shown for each redshift bin.  At $z>2$, we use the boundary condition defined at $1<z<2$ since \citet{williams2009} provide selection criteria up to those redshifts.  We also slightly modify the diagonal boundary at $0.5<z<1$ to better accommodate the COSMOS data using the following relation: $U-V>1.08 \times (V-J)+0.43$.  Small modifications in the $UVJ$ selection such as this are not uncommon \citep[e.g.,][]{whitaker2011} and possibly reflect variations in the observed filter set and SED templates used to derive the rest-frame colors.  Based on the $UVJ$ selection, Figure~\ref{fig_cumnum_qg_frac} shows the fraction of QGs and SFGs as a function of redshift.  The error bars in Figure~\ref{fig_cumnum_qg_frac} at $z>1$ are larger due to the smaller sample size of the UDS data.  Below $z<1$, the wide area COSMOS data allow us to better constrain the properties of the most massive galaxies.  The QG fraction increases from $\sim 23\%$ at $z \sim 2.75$ to $\sim 89\%$ at $z \sim 0.375$.  We note that our quiescent and star-forming fractions are in good agreement with those of \citet{brammer2011} who reported on galaxies to $z \sim 2$ (open symbols in Figure~\ref{fig_cumnum_qg_frac}).  

Together, Figures~\ref{fig_uvj_ncum} and \ref{fig_cumnum_qg_frac} show that the progenitors of nearby massive galaxies at $z \sim 3$ were likely to be star forming given that their colors coincide with those of SFGs in $UVJ$ color space.  From $z \sim 3$ to $z \sim 1.5$ there appears to have been a substantial buildup in the population of QGs, which become the dominant population at $z \lesssim 1.5$ for galaxies selected at $n_{\rm c}$.  At $z<1.5$, the population of QGs continues to grow.  We examine how the structural properties have evolved in the next section.

\subsection{Structural Evolution}

\begin{figure*}
\epsscale{1.2}
\plotone{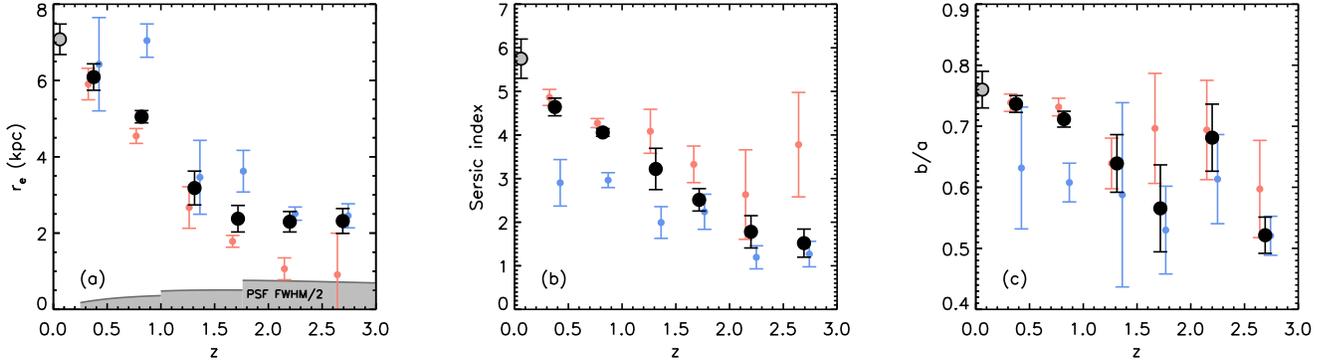}
\caption{Evolution in the structural properties of galaxies selected at a constant cumulative number density of \ncum. (a) Effective radius vs. redshift.  The shaded region indicates where $r_e <$~FWHM/2. (b) S\'{e}rsic index vs. redshift.  (c) Axis ratio vs. redshift.  The black circles represent the median for the full $n_{\rm c}$ selected sample while the red and blue data points represent the median values for the QG and SFG sub-populations, respectively (offset in redshift for clarity).  The gray filled circles at $z=0.06$ represent the median values for an SDSS sample at $z=0.06$ from \citet{szomoru2013}.  The typical size of a galaxy at $n_{\rm c}$ increases by a factor of $\sim 3-4$ since $z \sim 3$, with most of this change occurring at $z<2$.  The apparent constant size at $1.5<z<3$ is a consequence of the changing mix of QGs and SFGs combined with the overall growth of galaxy masses and sizes over this redshift range.  The S\'{e}rsic indices increase from $n \sim 1$ at high redshift to $n \sim 6$ at low redshift.  This suggests that most of the stars in galaxies at $n_{\rm c}$ were distributed in a disk at $z \sim 3$, while at low redshift they are distributed in a bulge.  The increasing axis ratios toward low redshift further support this view.  The median axis ratio of $b/a \sim 0.52$ at $z \sim 2.75$ is close to what is expected for randomly oriented thin disks, while at $z \sim 0$ the axis ratios are more indicative of spheroidal systems.} \label{fig_cumnum_structure}
\end{figure*}

\begin{figure*}
\epsscale{1.2}
\plotone{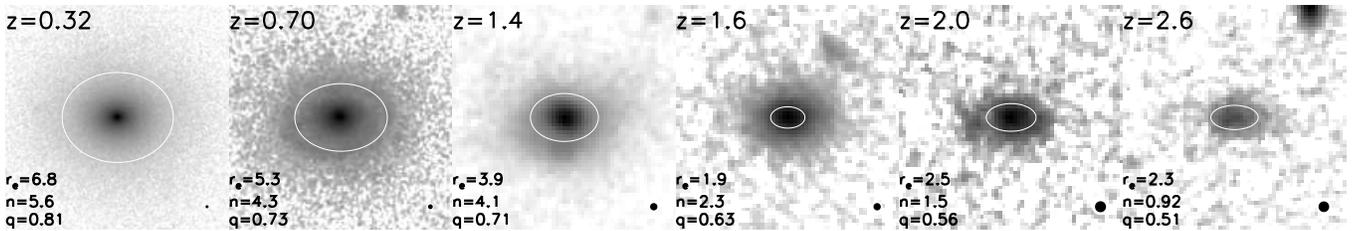}
\caption{Example postage stamps for galaxies selected at a constant cumulative number density of \ncum\ at different redshifts.  At a given redshift, objects were selected to have properties that follow the general trends seen in Figure~\ref{fig_cumnum_structure}.  Each postage stamp is $30$~kpc on a side and is rotated such that the major axis is aligned horizontally.  The PSF FWHM is indicated by the circle in the bottom right of each panel and the effective radius (in kpc), S\'{e}rsic index, and axis ratio ($q$) are given in the bottom left.  The half-light ellipse (shown in white) grows larger toward lower redshift indicating that more light is added to the outer parts, leading to the larger sizes, S\'{e}rsic indices, and the buildup of stellar mass.  Toward higher redshifts, the median axis ratio declines, suggestive of randomly oriented disks.} \label{fig_psncum}
\end{figure*}

\begin{figure*}
\epsscale{1.2}
\plotone{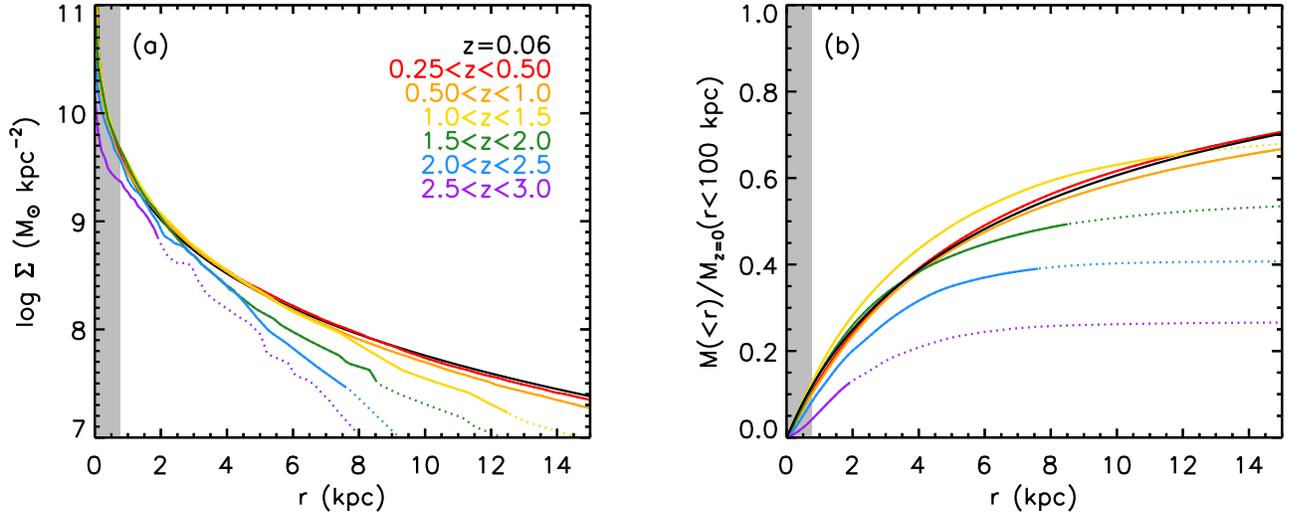}
\caption{(a) Stellar mass surface density profiles of galaxies selected at a constant cumulative number density of \ncum\ for different redshifts.  At a given redshift, light profiles were derived for each galaxy in the 0.3~dex mass bin based on the best fitting S\'{e}rsic index and effective radius.  These light profiles were then normalized to the stellar mass of each galaxy and then median combined.  The dotted portion of each profile indicates where the bootstrapped uncertainty of the median is greater than $20\%$.  The gray shaded region extends to the radius that corresponds to the maximum PSF FWHM/2 for the full sample (occurs at $z=1.76$).  (b) Cumulative stellar mass at a given radius relative to the total mass within $r<100$~kpc for a galaxy at $z \sim 0$.  The mass profiles overlap at small radii suggesting very little mass growth in the inner parts of a galaxy at $n_{\rm c}$, while at larger radii there appears to be a substantial buildup of mass with cosmic time.} \label{fig_cumnum_mass_profile}
\end{figure*}

\begin{figure}
\epsscale{1.2}
\plotone{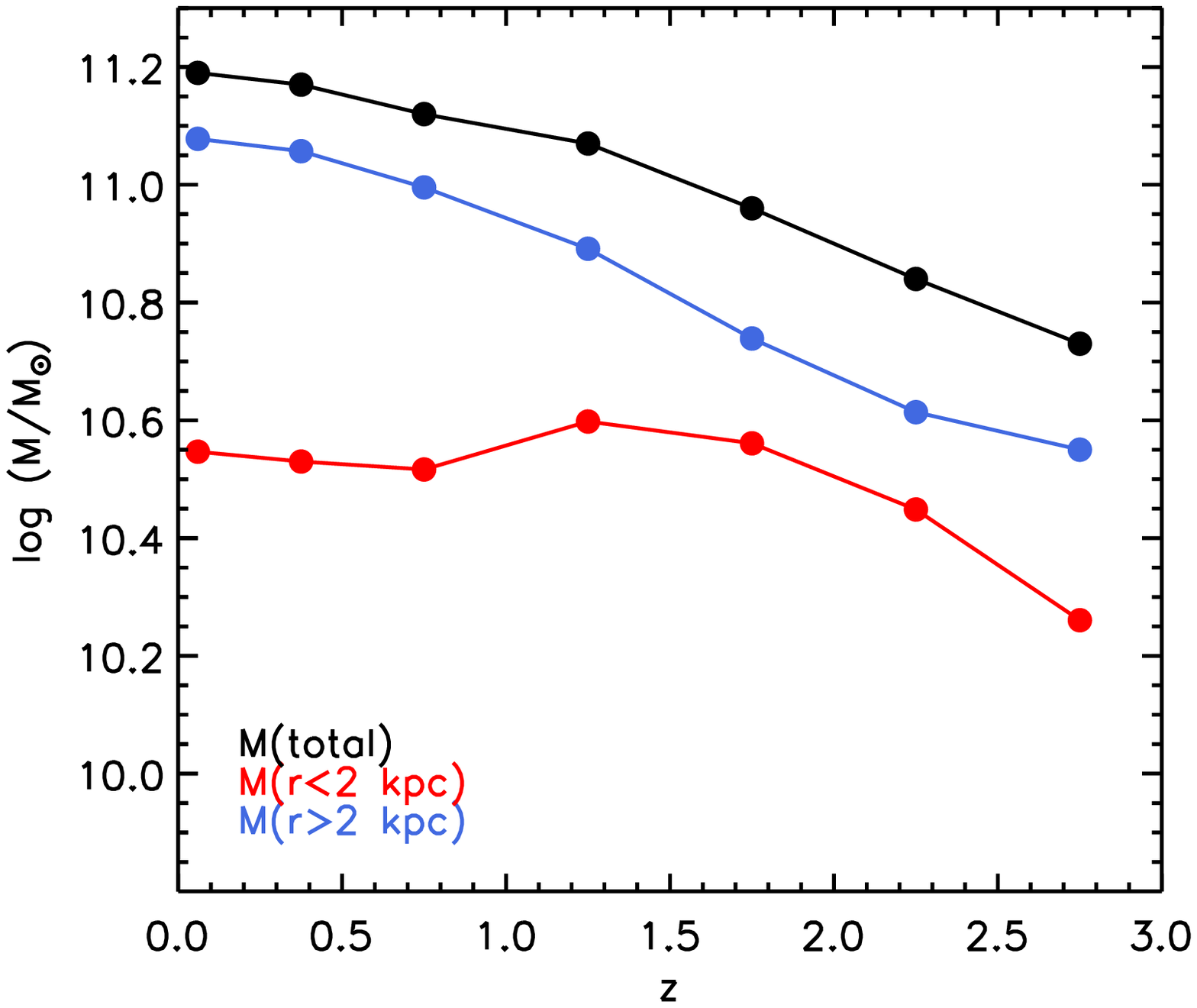}
\caption{Projected stellar mass for different radial regions of galaxies selected at a constant cumulative number density of \ncum.  At small radii ($r<2$~kpc, red line), most of the stellar mass was in place by $z \sim 2$.  At larger radii ($r>2$~kpc, blue line), there has been a substantial buildup of mass, fueling the overall mass growth of the galaxy.} \label{fig_cumnum_mass_rinout}
\end{figure}

We now turn to the structural properties of massive galaxies as they grow in time.  The general properties of galaxies selected at $n_{\rm c}$ are summarized in Table~\ref{table_ncum}.  Figure~\ref{fig_cumnum_structure} shows the evolution of the median effective radius, S\'{e}rsic index, and axis ratio for galaxies at \ncum.  We include in these figures a sample of SDSS galaxies at $z=0.06$ from \citet{szomoru2013} selected at the appropriate mass from Equation~(\ref{eq_mass_redshift}).  The SDSS data is not used in any of the fits that follow.  Below $z<2$, where QGs become the dominant population in the sample, Figure~\ref{fig_cumnum_structure}(a) shows that the effective radius increases substantially from $r_e \sim 2$~kpc at $z \sim 2$ to $r_e \sim 7$~kpc at $z \sim 0$.  The size evolution at $0.25<z<2$ follows

\begin{eqnarray}
r_e=(9.3 \pm 1.0)~{\rm kpc} \times (1+z)^{-1.1 \pm 0.2}
\end{eqnarray}
with the exponent being consistent with the value of $-1.27$ found in \citet{vandokkum2010} over roughly the same redshift range.  A striking feature in Figure~\ref{fig_cumnum_structure}(a) is the lack of evolution at $1.5<z<3$ in the median effective radius.  We investigate this further by showing the evolution of QGs and SFGs separately in Figure~\ref{fig_cumnum_structure}.  The constant median $r_e$ arises because SFGs are larger than QGs and their relative abundance changes as a function of redshift.  Above $z>3$, the size evolution is likely determined almost solely by SFGs since they become an overwhelming majority of the population.  We therefore expect the sizes of galaxies to decrease above $z \gtrsim 3$ \citep[see, e.g.,][]{oesch2010b,mosleh2012} for samples selected at $n_{\rm c}$.   We can test whether the apparent constant value of $r_e$ at $1.5<z<3$ is a generic feature or a consequence of the particular value of $n_{\rm c}$ selected for our study.  At lower values of $n_{\rm c}$ (i.e., higher masses at a given redshift), we find that $r_e$ can increase gradually from $z \sim 3$ to $z \sim 2$.

The S\'{e}rsic index determines the distribution of light and hints at the presence of a bulge or disk.  The median S\'{e}rsic index in Figure~\ref{fig_cumnum_structure}(b) increases from $n \sim 1$ at $z=2.75$ to $n \sim 6$ at $z \sim 0$.  The S\'{e}rsic index evolution at \zwindow\ can be characterized by

\begin{eqnarray}
n=(6.7 \pm 0.5) \times (1+z)^{-0.9 \pm 0.1},
\end{eqnarray}
which is consistent with the exponent of $-0.95$ found in \citet{vandokkum2010}.  The S\'{e}rsic index evolution indicates that while most of the stars in $\sim 2M^{\star}$ galaxies in the nearby universe are distributed in a bulge, the stars in their progenitor galaxies at $z>2$ were distributed in structures resembling exponential disks.  We note that \citet{wuyts2011b} also find that SFGs at high redshift, which represent the majority at $n_{\rm c}$, generally have low S\'{e}rsic indices around $n \sim 1$.  The fact that QGs at $2<z<3$ in Figure~\ref{fig_uvj_ncum} generally have higher S\'{e}rsic indices than SFGs \citep[see also,][]{bell2012}, as is also the case at lower redshifts, further suggests that our S\'{e}rsic profile fitting measurements are not significantly biased by the limiting depth of the {\em HST} imaging for higher redshift galaxies.  In the Appendix, we show this to also be the case with a much larger, stellar mass limited sample (Figure~\ref{fig_uvj}).

While the S\'{e}rsic indices can be suggestive of a bulge or disk component, the axis ratio distribution provides a better constraint on the shapes of galaxies.  Owing to the high resolution of the {\em HST} imaging, we can examine the axis ratios of galaxies selected at $n_{\rm c}$ to $z \sim 3$.  The median axis ratio of galaxies at \ncum\ has increased significantly since $z \sim 3$.  At $z=2.75$, the typical axis ratio is $b/a \sim 0.52$, a low value that is indicative of a distribution of randomly oriented thin disks.  Meanwhile, at $z=0.06$, the axis ratio is $b/a \sim 0.76$, closer to what is expected for elliptical galaxies.  This value is in good agreement with SDSS studies of massive QGs at $z=0.06$ \citep{vanderwel2009b,holden2012}.  At the highest redshifts ($2.5<z<3$), the residuals to the single component S\'{e}rsic profile fits are smooth and visual inspection of these residuals suggests that the lower axis ratios at high redshift are not generally driven by multiple components.  In addition, varying the number density selection to lower values, such as $n=10^{-4}$~Mpc$^{-3}$, does not impact the general decreasing trend of the median axis ratios towards higher redshifts.  Finally, as seen in Figure~\ref{fig_cumnum_structure}(c) the axis ratios of SFGs are generally lower than that of QGs at a given redshift and SFGs increasingly become the dominant population at $z>1.5$.  The axis ratios therefore also indicate, in addition to the S\'{e}rsic indices, that the stars in the progenitors of $\sim 2M^{\star}$ galaxies were distributed in disks at $z \sim 3$.

Figure~\ref{fig_psncum} shows example postage stamps of galaxies selected at $n_{\rm c}$ at different redshifts.  For illustrative purposes, we selected galaxies with structural parameters that follow the general trends seen in Figure~\ref{fig_cumnum_structure} but with more consideration for the trend in axis ratios.  Each postage stamp is $30$~kpc on a side and note that the size of the galaxy in each redshift bin is larger than the PSF.  The relative sizes between redshift bins are therefore easy to compare in this figure when paired with the indicated half-light ellipses.  Toward lower redshifts, especially at $z<2$, Figure~\ref{fig_psncum} shows how light is added to the outer parts, leading to the increasing size and S\'{e}rsic index of galaxies selected at $n_{\rm c}$.  At higher redshifts, the axis ratios decrease, suggesting a larger contribution from disks.

Synthesizing the structural information above with the star formation activity discussed in Section~\ref{sec_cumnum_uvj}, we see that the progenitors of $\sim 2M^{\star}$ galaxies at $z \sim 3$ were star-forming disks with $r_e \sim 2$~kpc.  By $z \sim 1.5$, many of these star-forming disks disappear and give rise to a population of compact QGs.  By $z \sim 0$, these compact QGs evolved into the large-sized, bulge dominated, quiescent $\sim 2M^{\star}$ galaxies.

\subsection{Mass Assembly}

\begin{figure}
\epsscale{1.2}
\plotone{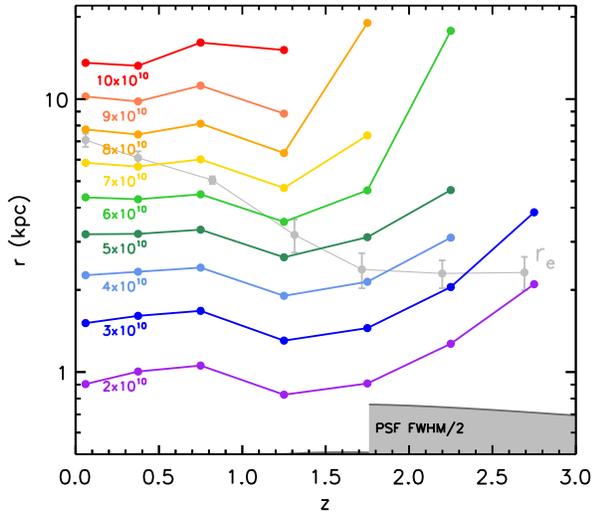}
\caption{Radius enclosing a fixed stellar mass as a function of redshift for a galaxy at \ncum.  Different colored lines indicate the evolution in the radius that encloses the given mass, ranging from $2 \times 10^{10}$ to $10^{11}$~\msun.  The lines therefore represent horizontal cuts in Figure~\ref{fig_cumnum_mass_profile}(b) (ignoring a normalization factor).  The lines of constant mass terminate in high-redshift bins where the given mass had not yet been assembled.  For reference, the effective radius for galaxies at fixed $n_{\rm c}$ is indicated by the gray line.  Below $z<2$, the radius enclosing a given mass remains roughly constant, indicating that new stellar mass growth toward low redshift occurs at larger radii.  From $z \sim 3$ to $z \sim 2$, the radius enclosing a given mass decreases but this trend may not be robust.} \label{fig_cumnum_radii_fixedmass}
\end{figure}

A more intuitive view of how the change in structural properties has impacted the evolution of galaxies at \ncum\ is given by the evolution of the mass surface density profiles in Figure~\ref{fig_cumnum_mass_profile}(a).  In order to compute these profiles, we first used the best fitting S\'{e}rsic index and effective radius for each galaxy to determine the light profile using the standard formula for a S\'{e}rsic profile:

\begin{eqnarray}
\Sigma(r) = \Sigma_e \exp(-b_n [(r/r_e)^{1/n}-1]), \label{eq_sersic}
\end{eqnarray}
where $\Sigma(r)$ is the surface brightness at radius $r$, $r_e$ the half-light radius, $\Sigma_e$ the surface brightness at $r_e$, $n$ the S\'{e}rsic index, and $b_n$ a constant that depends on $n$.  \citet{szomoru2012} find that the surface brightness profiles of galaxies at high redshift generally follow S\'{e}rsic profiles quite well.  For simplicity, we therefore use the analytic representation of the profile for each galaxy.  These light profiles were converted into mass profiles by normalizing the integrated light in the S\'{e}rsic profile to the stellar mass of each galaxy.  Note that this conversion neglects radial gradients in the mass-to-light ratio.  For a given redshift bin, the median of the mass profiles was computed at each radius resulting in the profiles shown in Figure~\ref{fig_cumnum_mass_profile}(a).  The uncertainty in the median of the mass profile at a given radius was computed by bootstrapping the sample.  We note that one-dimensional S\'{e}rsic fits to these median profiles recover the S\'{e}rsic index and half-light radius to within $\sim 10\%$ of the median values reported in Table~\ref{table_ncum}.

A naive interpretation of Figure~\ref{fig_cumnum_structure}(b) would be that the bulges of galaxies grow in time given the increase in the S\'{e}rsic index, a crude proxy for the bulge-to-disk ratio \citep[e.g.,][]{lackner2012}.  However, the mass profiles in Figure~\ref{fig_cumnum_mass_profile}(a) generally overlap at small radii and diverge at large radii, suggesting a buildup of mass in the outer parts of the galaxy with time.  Figure~\ref{fig_cumnum_mass_profile}(b) shows the cumulative proportion of mass assembled at different radii relative to the total mass within $r<100$~kpc of the median galaxy at $z=0.06$.  Roughly $\sim 50\%$ of the total mass of the galaxy is assembled within $r<7$~kpc at $z=0.06$, as expected given that $r_e \sim 7$~kpc at that redshift.  At $z \sim 2.25$, the assembled mass within $r<7$~kpc is $\sim 40\%$ of the total mass at $z=0.06$ indicating that much of the mass within $r<7$~kpc was already in place $\sim 10$~Gyr ago.  Note that the small sample in the $1<z<1.5$ bin likely leads to this curve falling slightly above the $0<z<1$ data at $r<10$~kpc.

In Figure~\ref{fig_cumnum_mass_rinout}, we compare the mass growth between the central and outer regions of galaxies selected at \ncum.  The total stellar mass as a function of redshift is shown by the black line, while the projected mass inside and outside of $r=2$~kpc is given by the red and blue lines, respectively.  These values are determined by integrating Equation~(\ref{eq_sersic}) as follows:

\begin{eqnarray}
M(r_{\rm in}<r<r_{\rm out}) = \int_{r_{\rm in}}^{r_{\rm out}} \Sigma(r)2\pi r dr
\end{eqnarray}
where $r_{\rm in}$ and $r_{\rm out}$ are the inner and outer radii enclosing the mass, $M$.  For the central regions ($r_{\rm in}=0$~kpc, $r_{\rm out}=2$~kpc), the stellar mass appears to grow from $z \sim 3$ to $z \sim 2$ but then levels off around $\sim 10^{10.5-10.6}$~\msun.  In contrast, in the outer regions ($r_{\rm in}=2$~kpc, $r_{\rm out}=100$~kpc) mass continues to build up over the entire redshift range studied, growing by a factor of $\sim 3$.  The stellar mass that has been added to the outer parts of galaxies over time is therefore the dominant source of assembled mass, as the central parts appear to have been assembled by $z \sim 2$.  Increasing the value of $r_{\rm in}$ so as to avoid the central regions of the S\'{e}rsic fit (e.g., $r_{\rm in}=1$~kpc), which can be less secure, does not qualitatively impact the latter result.  The results above are in qualitative agreement with those of \citet{vandokkum2010}.

An alternative projection of Figure~\ref{fig_cumnum_mass_rinout} is shown in Figure~\ref{fig_cumnum_radii_fixedmass}.  This figure shows the radius enclosing a given stellar mass as a function of redshift \citep[analogous to Figure~1 in][]{diemand2007b} for galaxies at \ncum.  It is the equivalent of taking horizontal cuts in Figure~\ref{fig_cumnum_mass_profile}(b), and scaling by a stellar mass.  The figure therefore depicts the evolution in the ``onion layering'' of stellar mass.  The lines of constant mass terminate in high-redshift bins where the given mass had not yet been assembled.  Below $z<2$, the radius enclosing a given mass remains roughly constant.  This again highlights that new stellar mass growth below $z<2$ occurs at larger radii and that the mass within the inner parts remains roughly unchanged.  Between $z \sim 3$ and $z \sim 2$, the radius enclosing a given stellar mass appears to decrease by roughly a factor of $\sim 2$.

\section{Discussion} \label{sec_discussion}

\subsection{Inside-out Mass Growth}

In Section~\ref{sec_cumnum}, we have shown that as a galaxy grows in stellar mass by a factor of $\sim 3$ from $M \sim 5 \times 10^{10}$~\msun\ at $z=2.75$ to $M \sim 1.5 \times 10^{11}$~\msun\ at $z \sim 0$, most of the new stellar mass that is added contributes toward mass growth at larger radii as one moves toward lower redshifts.  The mass profiles presented in Figure~\ref{fig_cumnum_mass_profile} highlight this point as they show that most of the mass in the core is in place by $z \sim 2$.  Below $z \sim 2$, Figure~\ref{fig_cumnum_mass_rinout} shows how mass growth continues in the outer parts to the present epoch.  As a consequence, the effective radius of the galaxy grows from $r \sim 2$~kpc at $z \sim 2-3$ to $r \sim 7$~kpc at $z \sim 0$.  Fitting the mass and effective radius evolution at $z<2$, we find that $r_e \propto M^{2.0 \pm 0.3}$, which is almost exactly the relation found in \citet{vandokkum2010} at $z \lesssim 2$.

The high resolution of the {\em HST} imaging allowed us to probe the mass distribution of galaxies at $z>2$ at very small radii.  This was not possible in the \citet{vandokkum2010} study as it relied on ground-based imaging.  With the higher resolution {\em HST} imaging, we find that the mass within $r=2$~kpc, which is roughly the median effective radius for a galaxy selected at $n_{\rm c}$ at $z>2$, remains roughly constant at $z<2$.  Thus, there does not appear to be any substantial growth at $z<2$ in the central part of the bulge that will characterize this galaxy at $z \sim 0$.  From $z \sim 3$ to $z \sim 2$, there is evidence for mass growth in the inner part of the galaxy, as seen in Figures~\ref{fig_cumnum_mass_rinout} and \ref{fig_cumnum_radii_fixedmass}.  We note, however, that selecting galaxies at different values of $n_{\rm c}$ can result in a more constant enclosing radius across redshift for a given mass.  Larger samples at $z>2$ may help in clarifying the potential mass buildup in the inner parts of galaxies at those early times.

Dissipationless minor mergers are thought to play a significant role in the buildup of mass for QGs.  In addition, such processes are predicted to assemble mass at large radii, thereby contributing toward the size growth of galaxies \citep[e.g.,][]{bournaud2007c,naab2009b,bezanson2009}.  For example, using a cosmological hydrodynamical simulation, \citet{naab2009b} suggest that a $z \sim 0$ galaxy with $M \sim 1.5 \times 10^{11}$~\msun\ grew by a factor of $\sim 3$ after $z \sim 3$, primarily through dry minor mergers.  This mass growth is almost exactly what is found in this work for the same mass descendant at $z \sim 0$.  Over this 11~Gyr timespan, the galaxy grew in size by a factor of $\sim 3-4$, also in agreement with our results.  One caveat in this comparison is that the overall sizes of the simulated galaxy are a factor of $2-3$ lower than what is found in this work.  With more recent simulations, \citet{hilz2013} show that the relation $r_e \propto M^{2}$ \citep[see also,][]{laporte2012}, found in this work and in \citet{vandokkum2010}, is most easily reproduced by $\sim 3-5$ mergers with mass ratios of 1:5.  In contrast, major mergers lead to a close to linear dependence of $r_e$ on $M$, which is not supported by the observations.  It remains to be seen whether these mergers are actually observed \citep[see, e.g.,][]{williams2011,newman2012}.

\subsection{The Progenitors of Local Ellipticals at $z \sim 3$}

In the local universe, the most massive systems are generally bulge dominated elliptical galaxies \citep[e.g.,][]{vanderwel2009b,holden2012} with large effective radii \citep[e.g.,][]{shen2003}.  In this work and in \citet{vandokkum2010}, we showed that the progenitors of these local ellipticals at $1 \lesssim z \lesssim 2$, are primarily compact QGs.  As discussed in the previous section, these compact QGs are generally considered the cores of ellipticals at $z \sim 0$ \citep[see also,][]{hopkins2009j}, growing in mass and size potentially through dissipationless minor mergers to match the properties, namely sizes, of local ellipticals.

With the {\em HST} imaging and the deep-IR UDS data, our number density selection allows us to the trace the properties of progenitors of local ellipticals to $z \sim 3$.  At these redshifts, Figures~\ref{fig_uvj_ncum} and \ref{fig_cumnum_qg_frac} show that most of the progenitors are SFGs.  The S\'{e}rsic indices in Figure~\ref{fig_cumnum_structure}(b) suggest that these SFGs have exponential profiles, which is typically associated with the surface brightness profiles of disks.  The S\'{e}rsic index alone, however, is not definitive in defining the shapes of these progenitors at $z \sim 3$.  Instead, the most compelling evidence comes from the axis ratios shown in Figure~\ref{fig_cumnum_structure}(c).  At $z \sim 0$, galaxies selected at \ncum\ have median observed axis ratios of $b/a \sim 0.75$.  The implied intrinsic axis ratio is roughly 2:3 \citep{holden2012}, indicative of spheroidal systems.  Meanwhile, at $z \sim 3$, Figure~\ref{fig_cumnum_structure}(c) shows that galaxies at $n_{\rm c}$ have a median axis ratio of $b/a \sim 0.52$.  This value for the axis ratio is very low considering that for a population of randomly oriented, infinitely thin disks, the median axis ratio would be $b/a \sim 0.5$.  It is therefore likely that the progenitors of massive, quiescent, local elliptical galaxies at $z \sim 3$ are star-forming disks.  Many of the star-forming disks at $z \sim 3$ disappear over the redshift range $1.5<z<3$ and give way to compact QGs, as seen in Figure~\ref{fig_uvj_ncum}.  While the details of this transition require further investigation, we note that for a much broader mass range than what is considered here, \citet{barro2012} identify a subset of the star forming population at $2<z<3$ with similar structure to that of compact QGs \citep[see also,][]{stefanon2013}.

\section{Summary} \label{sec_summary}

We used {\em HST} imaging to study the structural properties of galaxies selected at a constant cumulative number density of \ncum\ at redshifts of \zwindow.  This selection allowed us to trace the evolution of galaxies with stellar mass $M=5 \times 10^{10}$~\msun\ at $z=2.75$, as they grew by a factor of $\sim 3$ to become $\sim 2M^{\star}$ galaxies in the local universe.  This work builds on the previous analysis by \citet{vandokkum2010}, who also selected galaxies at a constant number density.  Here, we employ high-resolution {\em HST} imaging and extend the analysis to $z \sim 3$.  In contrast to mass-selected studies, our selection at a constant cumulative number density allows for a more straightforward evolutionary link between progenitors and descendants at different redshifts.  At \zuds, we employed CANDELS WFC3 $J_{125}$ and $H_{160}$ imaging in the UDS, while at \zcosmos\ we used wide-field ACS $I_{814}$ imaging in COSMOS to fit single-component S\'{e}rsic profiles at a common rest-frame wavelength.  The resulting S\'{e}rsic indices, effective radii, and axis ratios were used to aid in our analysis.  The uniform data sets and analysis methods carried out in this work serve to minimize systematics.

Our main conclusions are the following:

\begin{enumerate}

\item The typical galaxy at the selected value of $n_{\rm c}$ has grown in effective radius by a factor of $\sim 3-4$, mostly since $z \sim 2$ (Figure~\ref{fig_cumnum_structure}).  

\item The evolution in the stellar mass surface density profiles of galaxies selected at $n_{\rm c}$ indicates that most of the mass in the central regions was in place by $z \sim 2$, while almost all of the new mass growth took place in the outer parts (Figures~\ref{fig_cumnum_mass_profile}-\ref{fig_cumnum_radii_fixedmass}).  This inside-out mass growth is responsible for the increase in size and S\'{e}rsic index toward low redshift.  

\item At $z<2$, we find that as the stellar mass builds up, the effective radius grows as $r_e \propto M^{2.0}$, in excellent agreement with \citet{vandokkum2010}.  Recent simulations show that such a dependence is consistent with mergers with mass ratios of 1:5 being responsible for much of the size growth \citep{hilz2013}.

\item At $z \sim 3$, the rest-frame $UVJ$ colors, S\'{e}rsic indices, and axis ratios indicate that the progenitors of present day massive galaxies were star-forming disks with $r_e \approx 2$~kpc and a third of the $z \sim 0$ stellar mass.  At $1.5 \lesssim z \lesssim 2$, these galaxies doubled in stellar mass and were mostly compact QGs.  These galaxies evolved further into the $\sim 2M^{\star}$ galaxies in the local universe that are known to be quiescent, bulge dominated, elliptical galaxies with large effective radii.

\end{enumerate}

\acknowledgments
We thank the anonymous referee for helpful comments and suggestions.  We also thank Daniel Szomoru for helpful discussions.  We acknowledge funding from ERC grant HIGHZ 227749.  This research was supported by an NWO-Spinoza Grant.




\clearpage
\begin{appendix}

\section{The Structural Properties of Quiescent and Star Forming Galaxies for Stellar Mass Limited Samples} \label{sec_fixedmass}

\begin{figure*}[b!]
\epsscale{1.2}
\plotone{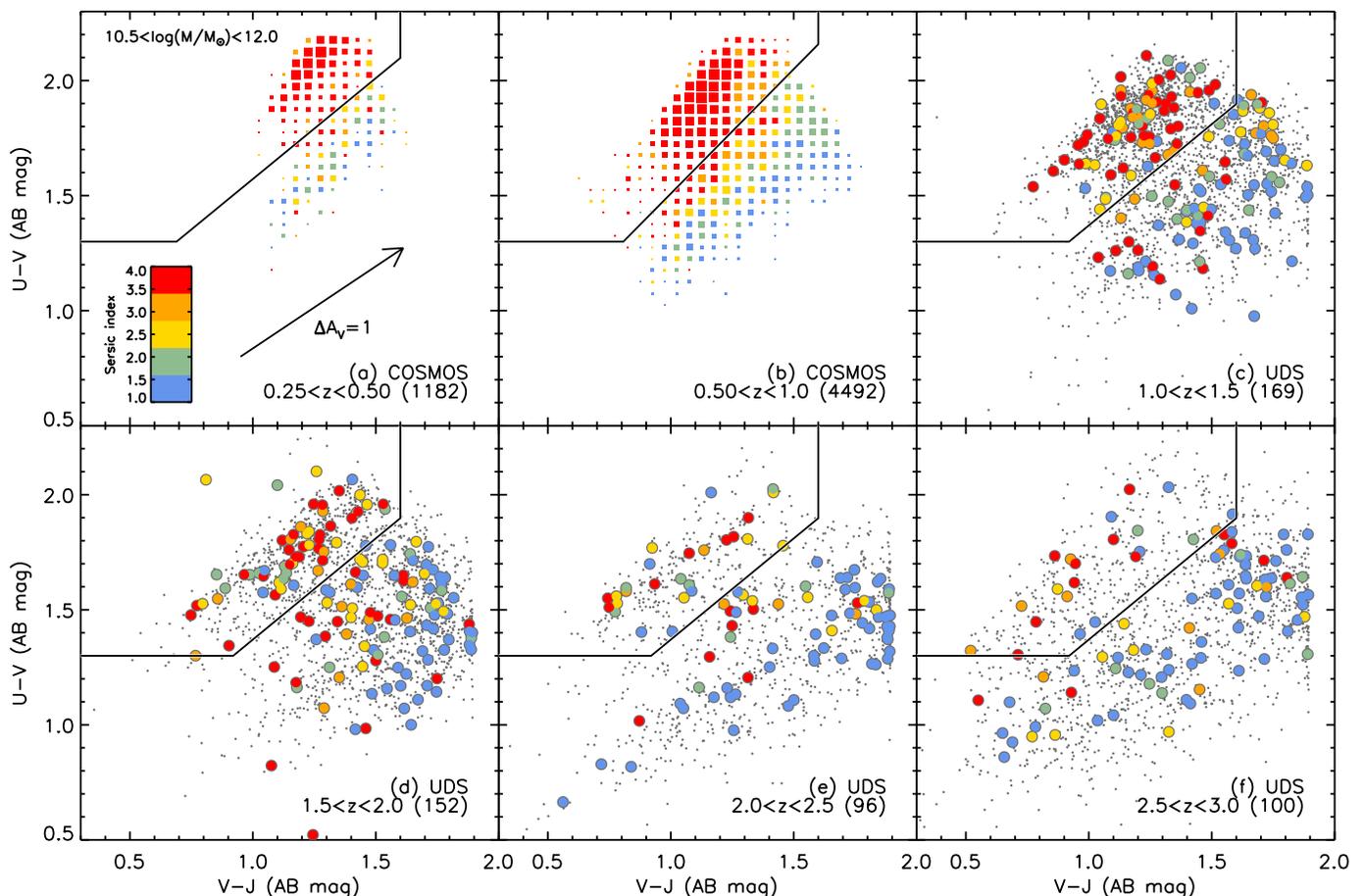}
\caption{Rest-frame $U-V$ vs. $V-J$ for galaxies with mass, \masslimit, in COSMOS and the UDS.  In COSMOS ($0.25<z<1$), each color-color bin is color-coded by the median S\'{e}rsic index and the symbol size is indicative of the number of galaxies in the bin (weighted on a logarithmic scale).  Bins with less than two galaxies are not color-coded.  In the UDS ($1<z<3$), the gray points indicate the parent sample of galaxies while objects color-coded by S\'{e}rsic index indicate the main subsample with {\em HST} imaging.  The sample size of galaxies with measured structural parameters is indicated in the bottom right of each panel.  The solid black line shows the division between QGs and SFGs.  At a given redshift, $UVJ$-selected QGs generally have higher S\'{e}rsic indices compared to SFGs.} \label{fig_uvj}
\end{figure*}

The primary purpose of the paper was to examine the structural properties of galaxies selected at a constant cumulative number density.  In doing so, we were able to trace the evolution of $\sim 2M^{\star}$ galaxies from their progenitors at $z \sim 3$.  Here we show results for galaxies selected above a constant mass limit, a more commonly used selection that provides an alternative view to the number density selection.  We distinguish QGs and SFGs with a $UVJ$-selection, as these two subpopulations are known to have differing properties.  As the mass selection allows for a larger sample, we can confirm some of the general findings in the main part of the paper where the sample was much smaller due to the number density selection.

\subsection{Classification of QGs and SFGs with $UVJ$ Selection} \label{sec_classification}

Figure~\ref{fig_uvj} shows $UVJ$ diagrams for galaxies in different redshift bins at \zwindow\ above a stellar mass of \masslimit.  Above this mass limit, the sample is complete for QGs and SFGs to $z \sim 2.5$ \citep{quadri2012}, thus the highest redshift bin at $2.5<z<3$ likely exhibits some incompleteness for QGs.  The subset of objects with measured structural parameters from the {\em HST} imaging is color-coded according to their S\'{e}rsic index.  For the COSMOS sample at \zcosmos, each color-color bin is color-coded according to the median S\'{e}rsic index for galaxies in that bin and the symbol size reflects the size of the sample within the bin (on a logarithmic scale).  The boundary distinguishing QGs from SFGs is shown for each redshift bin.

For the redshift range studied here, Figure~\ref{fig_uvj} shows the buildup in the proportion of QGs, down to a fixed mass limit, with time.  The figure also shows that QGs generally have higher S\'{e}rsic indices relative to SFGs at a given redshift.  This is also the case for optically red galaxies (e.g., $U-V \gtrsim 1.5$), where $UVJ$-selected QGs have higher S\'{e}rsic indices relative to reddened SFGs.  The $UVJ$ selection therefore works efficiently to distinguish galaxies to $z \sim 3$ based not only on their SFHs, but also their structural properties.  This was suggested in the analysis at constant number density, but is confirmed here with a much larger sample.  The fact that QGs and SFGs have different structural properties at the highest redshifts provides additional confidence in the structural parameters obtained from GALFIT: high and low S\'{e}rsic index galaxies are found where we think they should lie, in the quiescent and star-forming regions of $UVJ$ color space, respectively.

\subsection{S\'{e}rsic Indices}

\begin{figure}[b!]
\epsscale{0.7}
\plotone{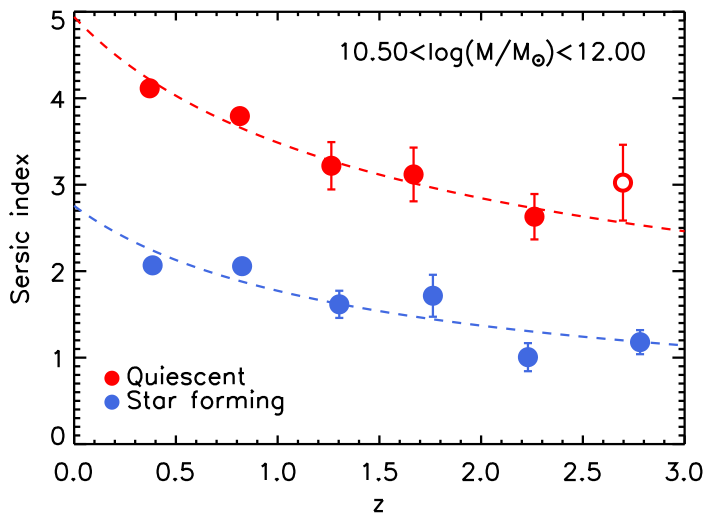}
\caption{S\'{e}rsic index vs. redshift for QGs (red) and SFGs (blue) with stellar mass \masslimit.  The $1\sigma$ error bars are computed from bootstrapping.  Dashed lines represent a fit to the data of the form $n \propto (1+z)^{\alpha}$, where $\alpha=-0.50 \pm 0.18$ and $-0.64 \pm 0.16$ for QGs and SFGs respectively.  For QGs, the highest redshift data point at $2.5<z<3$ (open circle) is not included in the fit due to incompleteness.  Because of potential systematic differences in the COSMOS (\zcosmos) and UDS (\zuds) samples, the fits were carried out with a systematic uncertainty of 10\% added in quadrature to the $1\sigma$ errors shown.  The S\'{e}rsic indices of both QGs and SFGs decrease toward higher redshift.} \label{fig_sersic}
\end{figure}

Figure~\ref{fig_sersic} shows the median S\'{e}rsic index, $n$, of QGs and SFGs at different redshifts.  Error bars ($\pm 1\sigma$) were computed by bootstrapping the sample in different redshift bins.  Note that the random errors at $z<1$ are generally smaller than the size of the data point given the large sample size in COSMOS.  QGs (red circles) show mild evolution in their S\'{e}rsic index over \zwindow, increasing slightly from $n \approx 3$ to $n \approx 4$ at low redshift.  A fit of the form $n \propto (1+z)^{\alpha}$ indicates that $\alpha = -0.50 \pm 0.18$.  We ignore the $z \sim 2.75$ QG data point for this fit as incompleteness likely plays a role for QGs at the highest redshifts.  In carrying out these fits, we also add a 10\% systematic uncertainty in quadrature to the random errors to account for potential differences between the COSMOS and UDS samples.  The median S\'{e}rsic index for SFGs (blue circles) increases from $n \approx 1$ at $2<z<3$ to $n \approx 2$ at low redshift.  A fit of the form $n \propto (1+z)^{\alpha}$ indicates that $\alpha = -0.64 \pm 0.16$.  The S\'{e}rsic indices of both QGs and SFGs with mass \masslimit\ are therefore decreasing toward higher redshifts.

A potential technical explanation for the declining S\'{e}rsic indices of QGs with redshift is the presence of SFGs which scatter into the QG bin.  The rest-frame colors are more uncertain at higher redshifts and there are relatively more SFGs, which generally have lower S\'{e}rsic indices.  We test this explanation by comparing the SSFRs of QGs with high S\'{e}rsic indices ($n>2.5$) with those that have low S\'{e}rsic indices ($n<2.5$).  Using the SSFRs from the SED fit we find that for the different redshift bins above $z>1.5$, the QGs with low and high S\'{e}rsic indices generally have median SSFRs within the uncertainties.  We note that for the highest redshift bin, $2.5<z<3$, the difference in SSFR is somewhat larger.  However, incompleteness for QGs likely plays a role for this redshift bin.  It is worth noting that for a similar mass limit as in this work, \citet{newman2012} also find that the S\'{e}rsic indices of QGs decline to $z=2.5$ with QGs selected based on an SSFR limit and a lack of detection in MIPS 24~\micron\ imaging.  Likewise for SFGs at low redshift, the more numerous QG population could scatter into the SFG selection, leading to an elevated median S\'{e}rsic index for SFGs.  The smaller uncertainties for the rest-frame colors make this scenario less likely.  We confirm that the typical SSFRs of SFGs at $0.25<z<1$ with low and high S\'{e}rsic indices are the same within the uncertainties.  We therefore conclude that the decline with redshift in the S\'{e}rsic index for QGs and SFGs with mass \masslimit\ is not a consequence of galaxies of either type scattering into the other bin.

Again, we note that at the highest redshifts, the S\'{e}rsic indices obtained from GALFIT are generally different for QGs and SFGs, as is the case at lower redshifts where the signal-to-noise of the measurements is more robust.  In addition to our simulations discussed earlier, this point affirms that our structural parameters are reasonable at high redshift.

\subsection{Sizes}

\begin{figure*}
\epsscale{1.15}
\plottwo{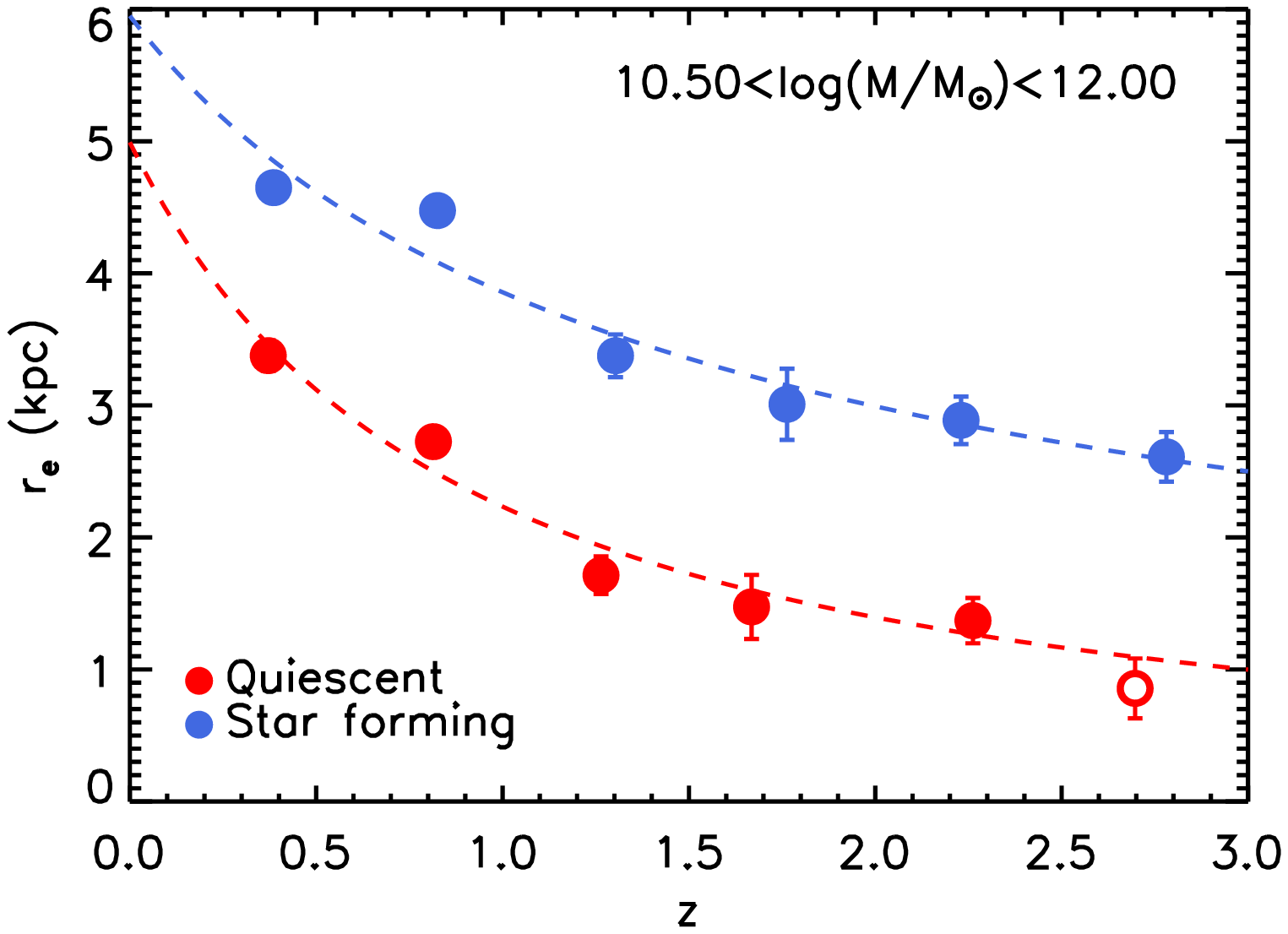}{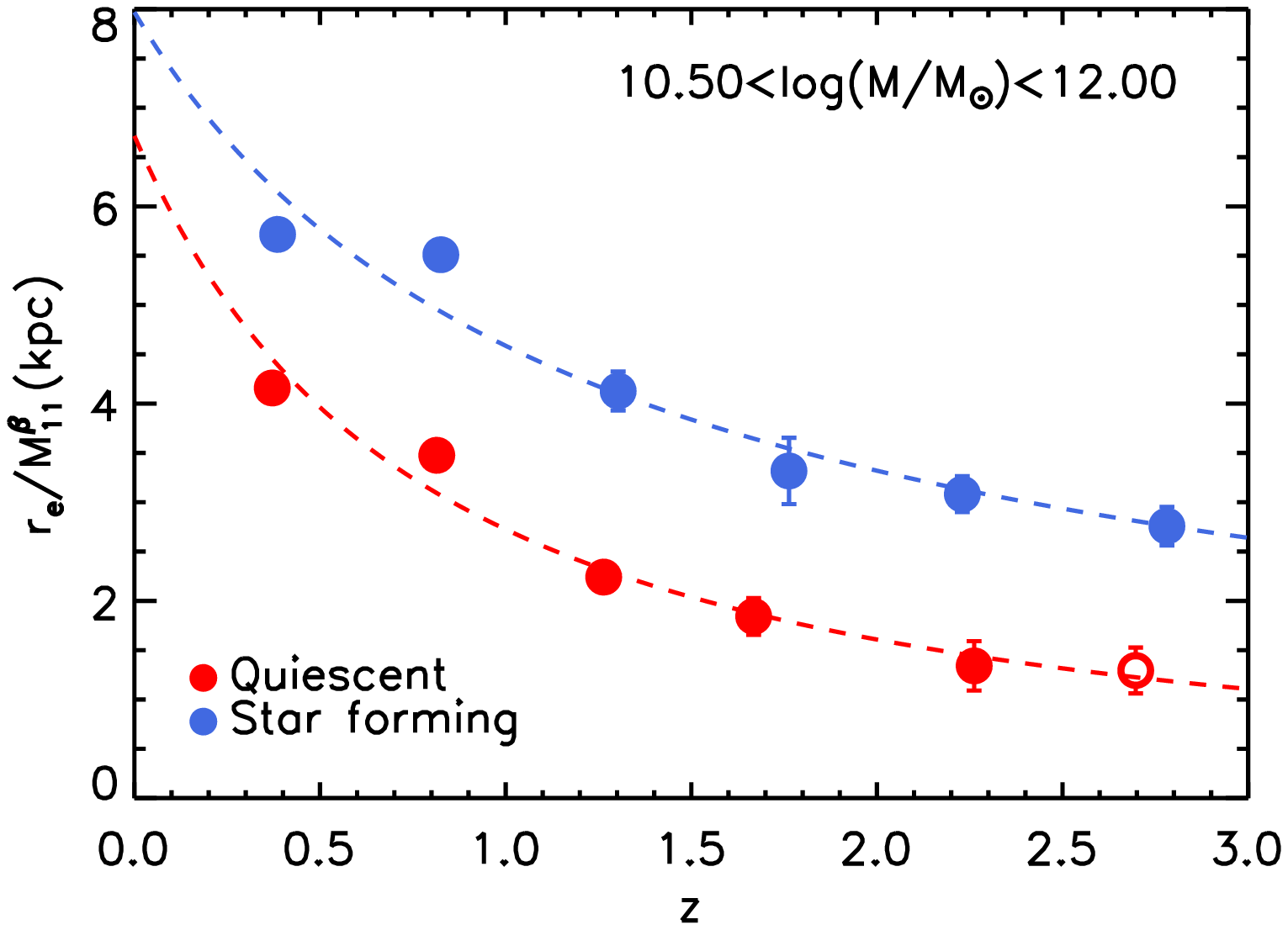}
\caption{(a) Effective radius vs. redshift for QGs (red) and SFGs (blue) with stellar mass \masslimit.  The $1\sigma$ error bars are computed from bootstrapping.  For a galaxy at a given redshift, the {\em HST} bandpass nearest to rest-frame $\sim 5200$~\AA\ was selected for the size measurement.  Dashed lines represent a fit to the data of the form $r_e \propto (1+z)^{\alpha}$, where $\alpha= -1.16 \pm 0.20$ and $-0.63 \pm 0.13$ for QGs and SFGs, respectively.  For QGs, the highest redshift data point at $2.5<z<3$ (open circle) is not included in the fit due to incompleteness.  The sizes of both QGs and SFGs decrease toward higher redshift. (b) Mass-normalized effective radius vs. redshift.  The effective radius of each galaxy has been scaled to the value for that of a $10^{11}$~\msun\ galaxy ($M_{11}=M/10^{11}$~\msun) assuming that $r_e \propto M^{\beta}$ at all redshifts, where $\beta=0.56$ for QGs and $\beta=0.3$ for SFGs \citep{shen2003}.  The redshift evolution for the mass-normalized radius is characterized by $\alpha=-1.30 \pm 0.20$ and $\alpha=-0.80 \pm 0.13$ for QGs and SFGs, respectively.} \label{fig_size}
\end{figure*}

Figure~\ref{fig_size}(a) shows the median effective radius ($r_e$) of QGs and SFGs at different redshifts.  QGs above \masslimit\ show substantial evolution in their sizes over \zwindow, increasing by roughly a factor of $\sim 3$ from $r_e \sim 1$~kpc to $r_e \sim 3$~kpc.  A fit of the form $r_e \propto (1+z)^{\alpha}$ indicates that $\alpha = -1.16 \pm 0.20$ for QGs.  The QG data point at $2.5<z<3$ was again not included in the fit due to incompleteness.  Meanwhile, SFGs increase in size by a factor of $\sim 2$ over the same redshift range from $r_e \sim 2.5$~kpc to $r_e \sim 4.5$~kpc.  We find $\alpha =  -0.63 \pm 0.13$ for SFGs.  We note that at high redshifts, the median sizes for QGs and SFGs are always larger than the measurement limits imposed by the PSF of the WFC3 $J_{125}$ (FWHM/2~$\sim 0\farcs06$) and $H_{160}$ (FWHM/2~$\sim 0\farcs09$) imaging. 

Figure~\ref{fig_size}(b) shows the median mass-normalized effective radius versus redshift for QGs (red) and SFGs (blue).  The effective radii were normalized to a stellar mass of $10^{11}$~\msun\ assuming a size-mass relation $r_e \propto M^{\beta}$, where $\beta=0.56$ for QGs and $\beta=0.3$ for SFGs \citep{shen2003}.  The mass-normalized radii therefore represent the size that galaxies would have if they lie on the given size-mass relation with a stellar mass of $10^{11}$~\msun.  A fit of the form $r_e/M_{11}^{\beta} \propto (1+z)^{\alpha}$, where $M_{11}=M/10^{11}$~\msun, indicates that $\alpha = -1.30 \pm 0.20$ for QGs and $\alpha=-0.80 \pm 0.13$ for SFGs.

For our mass-limited sample, the median size of QGs is generally smaller than that of SFGs at a given redshift and the sizes of both QGs and SFGs decrease toward higher redshift as also found by several other authors \citep[e.g.,][]{daddi2005,toft2007,cimatti2008b,franx2008,williams2010,newman2012}.

\subsection{Axis Ratios}

\begin{figure}
\epsscale{0.7}
\plotone{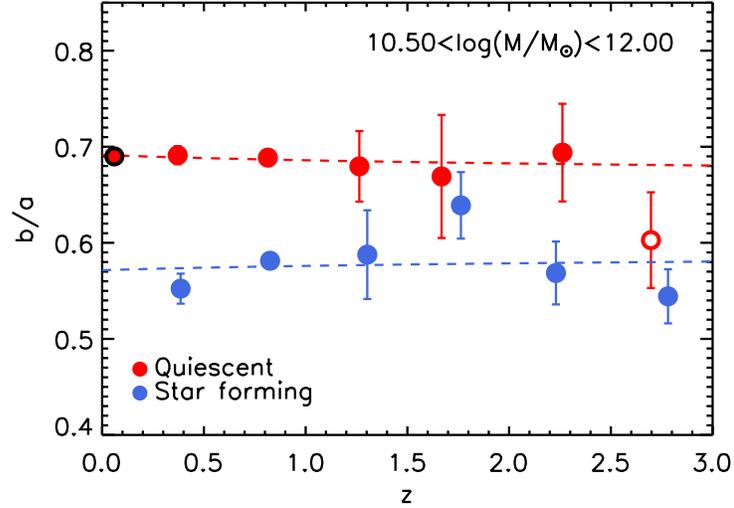}
\caption{Median axis ratio, $b/a$, vs. redshift for QGs (red) and SFGs (blue).  The $1\sigma$ error bars are computed from bootstrapping each subsample.  The filled circle at $z \sim 0$ represents the median $b/a$ value of QGs with mass $10.5<\log M/M_{\odot}<11$ computed by \citet{holden2012}.  Dashed lines indicate fits to our data of the form $b/a \propto (1+z)^{\alpha}$.  The QG data point at $2.5<z<3$ (open circle) was not included in the fit due to incompleteness in that bin.  For both QGs and SFGs, $\alpha$ is consistent with no evolution in the median axis ratio.} \label{fig_boa} 
\end{figure}

The distribution of axis ratios provides additional insight into the shapes of QGs and SFGs.  Figure~\ref{fig_boa} shows the median axis ratio of QGs and SFGs with mass \masslimit\ at different redshifts.  The dashed lines are fits of the form $b/a \propto (1+z)^{\alpha}$.  For QGs, a fit to the data at $0.25<z<2.5$ indicates $\alpha=-0.01 \pm 0.17$, consistent with no change over this redshift range in the median $b/a$.  The median axis ratio for all QGs at $0.25<z<2.5$ is $b/a \sim 0.69$, which is very close to the value computed by \citet{newman2012} for a similar mass limit (after accounting for IMF differences) and redshift range.  Incompleteness impacts the QG data point at $2.5<z<3$ and is therefore ignored in the fit above.  Also shown in Figure~\ref{fig_boa} is the median $b/a$ value at $z \sim 0$ from SDSS determined by \citet{holden2012} for galaxies with stellar mass $10^{10.5}<M/M_{\odot}<10^{11}$.  \citet{holden2012} find no change in the distribution of axis ratios of QGs from $z \sim 0$ to $z \sim 0.7$, consistent with our results at those redshifts.  For a sample of 14 QGs at $1.5<z<2.5$ with stellar masses $M>10^{10.8}$~\msun, \citet{vanderwel2011} computed a median axis ratio of $b/a=0.67$.  For the same redshift range and mass cut, we measure a median axis ratio of $b/a=0.73 \pm 0.06$ for 46 QGs, which is consistent with that of \citet{vanderwel2011}.  They suggest that QGs may be more disk dominated at high redshift based on the proportion of QGs with low axis ratios.  A more detailed analysis of the distribution of axis ratios with larger samples may lead to a more definitive answer.

The median axis ratio of SFGs also does not change significantly over \zwindow.  We find $\alpha=0.01 \pm 0.13$, consistent with no evolution.  The median value for the axis ratio of SFGs over \zwindow\ is $b/a \sim 0.58$.  The deviation at $1.5<z<2$ is within $\sim 1.8\sigma$ of this value.  Interestingly, for a population of infinitely thin disks, the median observed axis ratio would be $b/a \sim 0.5$, not far from the value found here.

The mass selection affords a larger sample at $2<z<3$ than was possible with the number density selection.  We find that the general trends found with the latter selection still hold with this much larger mass-limited sample.  In particular, the axis ratios of SFGs, which dominate both the mass-limited sample and the number density selected sample at $z>2$, display median axis ratios consistent with randomly oriented disks.

\subsection{Stellar Mass Variations}

The implemented stellar mass limit, \masslimit, to first order limits variations in the distribution of stellar masses across different redshift bins.  This works to minimize residual correlations between various parameters and stellar mass.  For QGs, the range of median stellar masses for the different redshift bins is only $0.07$~dex.  For SFGs, the range in median stellar masses is somewhat higher at $0.18$~dex, with the higher redshift bins ($z>1.5$) having more massive SFGs above the mass limit.  We do not expect this difference for SFGs to have a significant impact on our conclusions.  In fact, it strengthens many of our key points.  For example, if higher mass SFGs have higher S\'{e}rsic indices, then the S\'{e}rsic indices for SFGs at $1.5<z<3$ should be even lower than those at low redshift after correcting to the same median mass.

\end{appendix}

\end{document}